\documentclass[sigconf]{acmart}

\AtBeginDocument{%
  }

\copyrightyear{2025}
\acmYear{2025}
\setcopyright{rightsretained}
\acmConference[ASSETS '25]{The 27th International ACM SIGACCESS Conference on Computers and Accessibility}{October 26--29, 2025}{Denver, CO, USA}
\acmBooktitle{The 27th International ACM SIGACCESS Conference on Computers and Accessibility (ASSETS '25), October 26--29, 2025, Denver, CO, USA}
\acmDOI{10.1145/3663547.3759744}
\acmISBN{979-8-4007-0676-9/2025/10}

\begin{document}

\title{Speculating a Tactile Grammar: Toward Task-Aligned Chart Design for Non-Visual Perception}

\author{Areen Khalaila}
\affiliation{%
  \institution{Brandeis University}
  \city{Waltham}
  \state{MA}
  \country{USA}
}

\author{Dylan Cashman}
\affiliation{%
 \institution{Brandeis University}
 \city{Waltham}
 \state{MA}
 \country{USA}}


\begin{abstract}
Tactile graphics are often adapted from visual chart designs, yet many of these encodings do not translate effectively to non-visual exploration. Blind and low-vision (BLV) people employ a variety of physical strategies such as measuring lengths with fingers or scanning for texture differences to interpret tactile charts. These observations suggest an opportunity to move beyond direct visual translation and toward a tactile-first design approach. We outline a speculative tactile design framework that explores how data analysis tasks may align with tactile strategies and encoding choices. While this framework is not yet validated, it offers a lens for generating tactile-first chart designs and sets the stage for future empirical exploration. We present speculative mockups to illustrate how the Tactile Perceptual Grammar might guide the design of an accessible COVID-19 dashboard. This scenario illustrates how the grammar can guide encoding choices that better support comparison, trend detection, and proportion estimation in tactile formats. We conclude with design implications and a discussion of future validation through co-design and task-based evaluation.
\end{abstract}






\maketitle

\section{Introduction}

Tactile graphics serve as essential tools for accessing visual content in various domains, including education, healthcare, civic engagement, and personal data exploration \cite{Anqi, McDonald, Vandiver2019, jofre2016citizen}. For blind and low-vision (BLV) individuals, these graphics bridge a significant gap in access to visual information. However, tactile graphics are often designed by directly adapting visual charts without considering the perceptual and cognitive demands of tactile exploration \cite{zong2024umwelt}.

BLV users perceive tactile graphics through sequential and limited-resolution exploration. Techniques such as aligning fingers, counting ridges, distinguishing textures, and tracing edges are commonly used in reading tactile charts. These tactile strategies differ significantly from visual scanning behaviors and require dedicated design attention. Without an understanding of these strategies, designers risk producing charts that are technically correct but functionally unusable.

While previous work has offered guidelines for creating effective tactile graphics \cite{Fritz1999, Han2020, Jakub2022, Fan2020, Chen2025}, these tend to focus on specific chart types or encoding methods, and often assume a visual chart exists before accessibility is considered. What is missing is a flexible yet structured framework that starts from the user’s task and maps it to tactilely effective design strategies. Our framework is grounded in tactile perception, where information is gathered sequentially and through distinct strategies not present in visual scanning. These perceptual differences fundamentally alter how tasks map to effective encodings, enabling tactile-first designs rather than retrofits of visual charts

\section{Background and Related Work}
\subsection{Tactile Perception and Exploration}

Tactile perception differs fundamentally from visual perception. The human tactile system has a lower spatial resolution than the visual system and requires physical interaction to explore an object \cite{kuroki2025passive}. As such, tactile information is gathered sequentially, and the perception of spatial relationships depends heavily on the layout, spacing, and tactile distinguishability of elements.

Xu et al.~\cite{xu2023lets} introduced a grammar for physicalized data, focusing on expressive tactile chart generation. While this grammar enhances the flexibility of tactile chart authoring, it does not emphasize perceptual fit or task-strategy alignment. Our work builds on these findings by emphasizing the perceptual and cognitive mechanisms involved in tactile chart reading.

\subsection{Task-Driven Design}

In visualization research, it is well-established that the effectiveness of a chart depends on the alignment between user tasks and data encoding methods. For tactile graphics, this alignment becomes even more critical due to the additional constraints imposed by tactile exploration. Tasks such as comparing values, identifying trends, and estimating proportions require different tactile features to support accurate interpretation.

% Our framework builds upon task taxonomies developed in the data visualization literature and adapts them to the context of tactile interaction. By doing so, we provide a structure that supports tactile-first thinking in design.

\section{Tactile Perceptual Framework}

Tactile graphics must do more than just translate visual encodings—they must be constructed from the ground up to align with the tactile strategies of BLV users. To support this need, we introduce SenseMap, a perceptually grounded framework that connects user tasks, tactile strategies, and chart encodings.

SenseMap consists of three interconnected components:

\begin{itemize}
  \item \textbf{Task}: The analytical goal the user is trying to achieve (e.g., comparison, ranking).
  \item \textbf{Strategy}: The tactile method employed to complete that task (e.g., aligning fingers, counting, sweeping).
  \item \textbf{Encoding}: The design feature that supports that strategy (e.g., height-aligned bars, distinct textures).
\end{itemize}

This task-strategy-encoding model encourages tactile-first thinking: starting from the cognitive goal, identifying common tactile behaviors, and then selecting chart features that support those behaviors. Unlike visual-first retrofits, this model prioritizes perceptual accessibility from the outset.

We summarize the current structure of SenseMap in Table~\ref{tab:framework}.

\begin{table}[h]
\centering
\caption{Task--Strategy--Encoding Mapping}
\label{tab:framework}
\begin{tabular}{|p{1.25cm}|p{1.25cm}|p{4cm}|}
\hline
\textbf{Task} & \textbf{Tactile Strategy} & \textbf{Recommended Encoding} \\
\hline
Compare values & Aligning fingers & Equal-width bars with aligned tops for direct height comparison \\
\hline
Identify trends & Sequential sweeping & Raised line graphs with tactile anchor points at regular intervals \\
\hline
Estimate proportion & Segment estimation & Stacked bars with distinct textures for each segment and consistent tactile spacing \\
\hline
Rank categories & Counting and spacing & Discrete bars arranged with consistent gaps and textured increments \\
\hline
Detect outliers & Range scanning & Continuous lines with elevated tactile deviations and contrasting textures \\
\hline
\end{tabular}
\end{table}

Each mapping is grounded in previous empirical findings and design toolkits. Khalaila et al.’s \cite{Areen} replication of Cleveland and McGill’s graphical perception experiments with BLV participants demonstrated clear links between specific tactile exploration strategies and performance outcomes. Their findings showed that strategies such as aligning fingers or sweeping sequentially were associated with higher accuracy for tasks like value comparison and trend identification, while inconsistent spacing or unclear textures impeded performance. These results reinforce the importance of pairing user tasks with perceptually supportive encodings, as outlined in SenseMap.
Xu et al.~\cite{xu2023lets} introduced a generative grammar to produce tactile charts, which emphasizes flexible composition, but does not address perceptual strategies directly. SenseMap complements such generative efforts by emphasizing the alignment between user goals, tactile behavior, and encoding selection.
By building this alignment into the grammar, we move toward more principled, accessible tactile chart design. SenseMap serves not as a fixed prescription but as a guide to support task-aligned, perceptually effective encoding choices in tactile visualization.

\section{Illustrative Scenario: Public Health Dashboard}

\begin{figure*}[h!]
\centering
\includegraphics[width=0.3\textwidth]{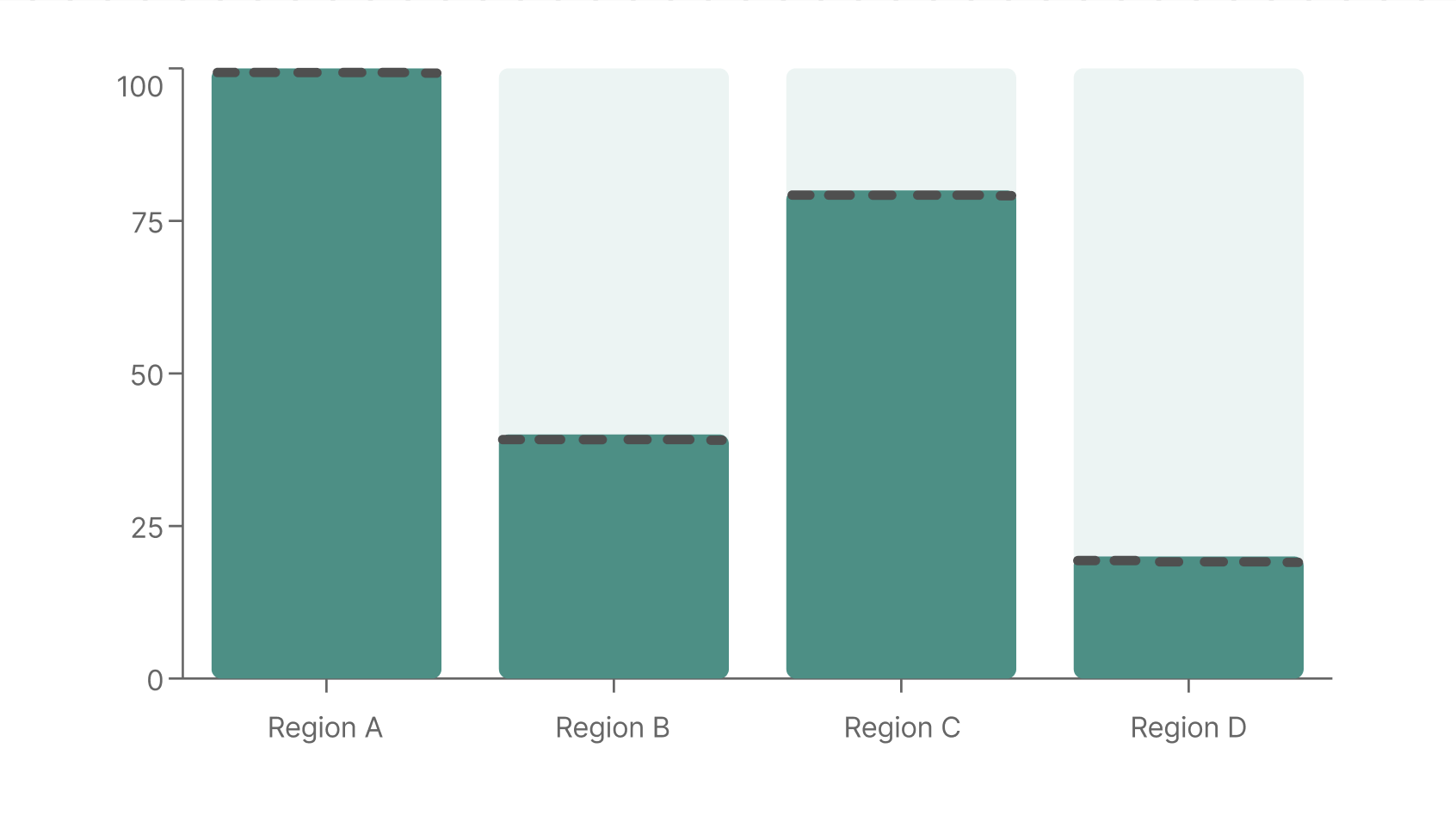}
\includegraphics[width=0.3\textwidth]{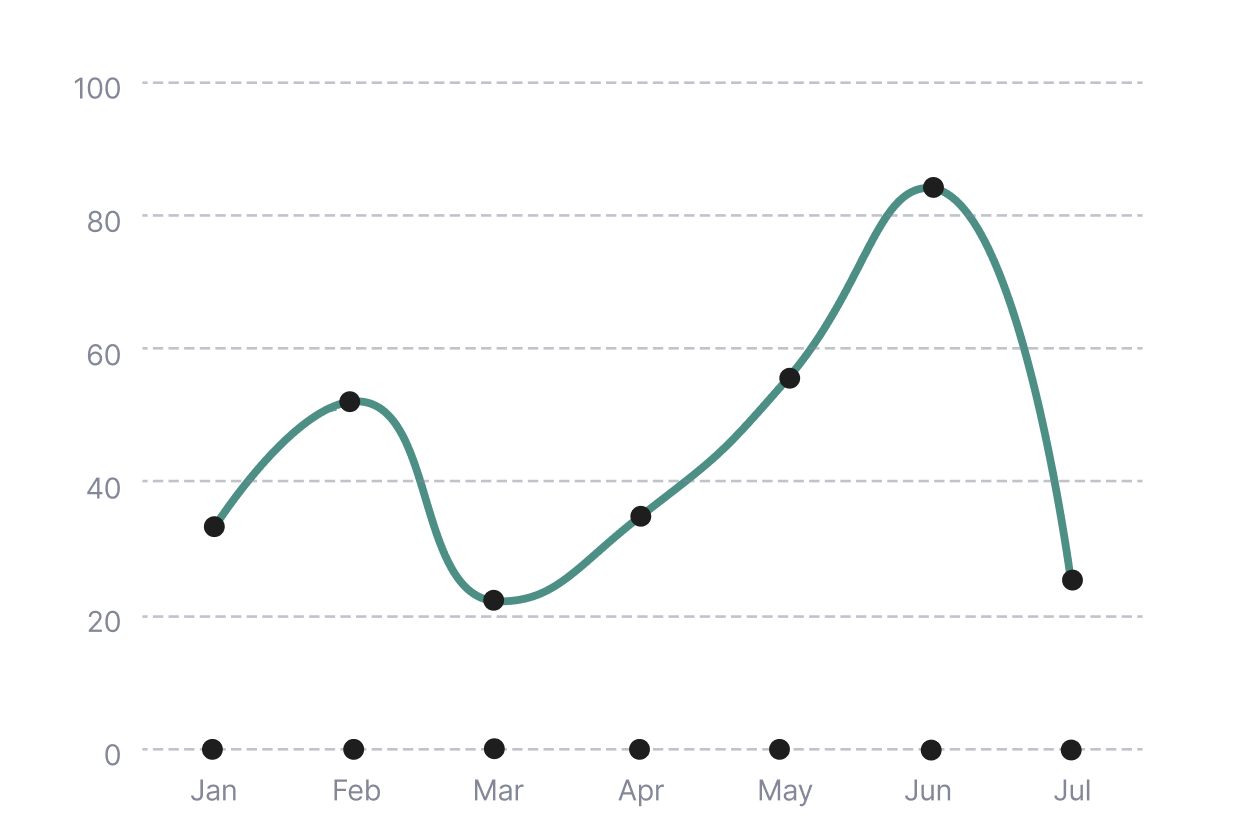}
\includegraphics[width=0.3\textwidth]{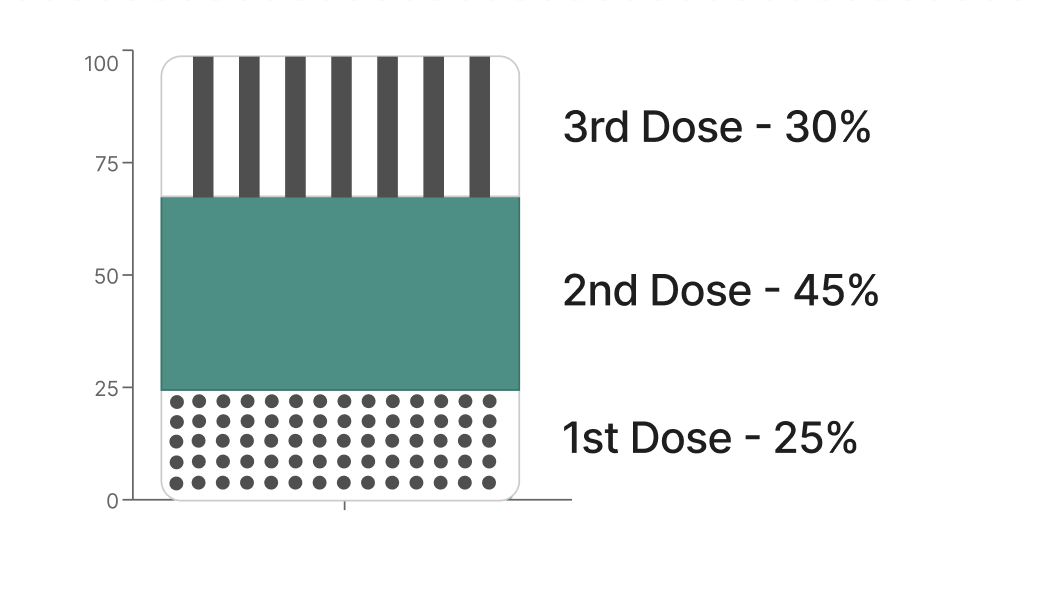}
\caption{Tactile chart mockups illustrating SenseMap. Left: Bar chart for infection rate comparisons (Task: Compare values, Strategy: Aligning fingers, Encoding: Equal-width bars with shared baseline). Center: Line chart for vaccination trend over time (Task: Identify trends, Strategy: Sequential sweeping, Encoding: Raised line with anchor points). Right: Stacked bar chart for dose coverage (Task: Estimate proportion, Strategy: Segment estimation, Encoding: Textured segments).}
\label{fig:dashboard-mockups}
\end{figure*}

To demonstrate the practical application of SenseMap, we describe how it could inform the design of a tactile COVID-19 public health dashboard. This example illustrates how different data tasks can be matched with perceptually aligned tactile encodings. Figure~\ref{fig:dashboard-mockups} presents three illustrative tactile charts designed using SenseMap.

\subsection{Infection Rates}
To compare infection rates across cities, the chart should use equal-width bars with a shared tactile baseline. By aligning finger positions across adjacent bars, users can directly compare relative heights. Uniform spacing ensures spatial consistency, allowing for repeatable measurements without visual reference. This chart design is shown on the left in Figure~\ref{fig:dashboard-mockups}.

\subsection{Vaccination Trend}
For representing vaccination uptake over time, a line chart with raised paths and tactile anchor points at each interval enables trend identification. The sequential sweeping strategy allows users to follow the line trajectory and detect increases or plateaus. Raised markers at data points help users track specific values and maintain orientation. This visualization appears in the center of Figure~\ref{fig:dashboard-mockups}.

\subsection{Dose Coverage}
To communicate vaccine dose distribution (e.g., first dose, second dose, third dose), a stacked bar encoding is appropriate. Each segment is differentiated by texture (e.g., dotted, ridged, smooth) to signify dose type. This supports a segment estimation strategy, allowing users to estimate the relative contribution of each dose type to the total. This chart appears on the right side of Figure~\ref{fig:dashboard-mockups}.

\subsection{Discussion}
This scenario highlights how the task-strategy-encoding mapping can drive effective tactile chart design. Rather than retrofitting a visual dashboard, SenseMap encourages a tactile-first approach, promoting usability, clarity, and data comprehension for BLV users.

Moreover, this example reveals a latent level of abstraction: distinct perceptual strategies and encoding features can be generalized across tasks and chart types. For instance, techniques like sequential sweeping or segment estimation transcend individual charts and represent reusable design patterns. This abstraction opens the door for a tactile grammar grounded not in visual analogs, but in the perceptual logic of touch.

\section{Design Guidelines and Implications}

Drawing from SenseMap, we propose the following guidelines to assist designers in creating perceptually grounded tactile charts:

\begin{itemize}
  \item \textbf{Begin with the Task}: Start by identifying the user's analytical goal—whether it’s comparison, trend analysis, or proportion estimation.
  \item \textbf{Support Natural Strategies}: Incorporate encodings that reflect tactile strategies such as finger alignment, texture differentiation, or spatial sweeping.
  \item \textbf{Prioritize Clarity}: Avoid visual clutter in tactile form. Use sufficient spacing, consistent alignment, and tactile contrast to make elements distinguishable.
  \item \textbf{Avoid Visual Bias}: Designs that work well visually—like pie charts or angle-based encodings—often perform poorly in tactile form.
  \item \textbf{Design Modularly}: Components of tactile encodings (e.g., bar heights, texture sets) should be reusable across contexts to promote scalable design.
\end{itemize}

These guidelines are meant to generalize across application domains, encouraging tactile-first thinking in both professional and educational settings. They can complement toolkits like Vysical~\cite{xu2023lets}, where generation flexibility could be paired with perceptual validation.

\section{Limitations and Future Work}

While SenseMap draws on strategies observed in BLV users, several limitations should be noted:

\begin{itemize}
  \item \textbf{Population Diversity}: We focus on users with total or partial vision loss. Users with deafblindness, cognitive disabilities, or reduced tactile sensitivity may require different strategies or encodings.
  \item \textbf{Static Medium}: SenseMap currently supports static tactile graphics. Interactive formats like refreshable displays or audio-tactile systems are not addressed.
  \item \textbf{Limited Task Set}: The task categories in SenseMap are representative but not exhaustive. More nuanced cognitive tasks could be incorporated.
\end{itemize}

To further formalize this framework, we plan to conduct a structured literature review of studies involving BLV participants and tactile charts. This review will identify the analytical tasks targeted (e.g., value comparison, trend identification, ranking), the tactile strategies described, and the encodings used. We will summarize these findings in a table mapping each task to its supporting literature and recommended design features, grounding our framework in published evidence.

Future directions include:
\begin{itemize}
  \item \textbf{Co-design with Diverse BLV Users}: Collaborating with users who have varying degrees of sensory ability to validate and expand the strategy set.
  \item \textbf{Tool Development}: Creating an authoring toolkit based on the framework to support designers in implementing perceptually aligned charts.
  \item \textbf{Empirical Evaluation}: Conducting user studies to compare framework-based designs with conventional visual-first tactile charts in task accuracy and user satisfaction.
  \item \textbf{Dynamic Charts}: Extending SenseMap to accommodate charts on refreshable displays and multimodal platforms.
\end{itemize}

\section{Conclusion}

We propose SenseMap, a tactile perceptual framework that maps user data analysis tasks to corresponding tactile strategies and effective encodings. By centering tactile perception rather than visual translation, SenseMap supports the creation of charts that are more usable and informative for BLV individuals. Rather than serving as a retrofit for charts originally designed for sighted audiences, SenseMap is intended to enable tactile-first designs that are conceived and optimized for non-visual perception, even in cases where no visual counterpart exists. This shift moves beyond adaptation toward creating representations grounded in tactile strategies and perceptual logic

We hope SenseMap will guide both researchers and practitioners in adopting a tactile-first mindset and spark new conversations about accessible data visualization design. Our future efforts will focus on validating SenseMap across broader user groups and integrating it into accessible design tools.

\bibliographystyle{ACM-Reference-Format}
\bibliography{sample-base}
\end{document}